\documentclass[12pt]{article}
\usepackage{amsmath,amssymb}
\usepackage{graphicx}
\usepackage{color}
\usepackage{cite,fancyhdr}
\pdfoutput=1

\usepackage[affil-it]{authblk}
\usepackage[left=2cm,right=2cm,top=3cm,bottom=3cm,a4paper]{geometry}

\begin{document}

\allowdisplaybreaks
\setcounter{footnote}{0}
\setcounter{figure}{0}
\setcounter{table}{0}

\title{\bf \large 
Nambu-Goldstone Boson Hypothesis for Squarks and Sleptons \\ in Pure Gravity Mediation
}
\author[1]{{\normalsize  Tsutomu T. Yanagida}}
\author[2]{{\normalsize Wen Yin}}
\author[2]{{\normalsize Norimi Yokozaki}}

\affil[1]{\small 
Kavli Institute for the Physics and Mathematics of the Universe (WPI),
 
University of Tokyo, Kashiwa 277--8583, Japan}

\affil[2]{\small 
Department of Physics, Tohoku University,  

Sendai, Miyagi 980-8578, Japan}

\date{}

\maketitle

\thispagestyle{fancy}
\rhead{ IPMU16-0123 \\ TU-1032 }
\cfoot{\thepage}
\renewcommand{\headrulewidth}{0pt}

\begin{abstract}
\noindent
We point out that a hypothesis of squarks and sleptons being Nambu-Goldstone (NG) bosons is consistent with pure gravity mediation or minimal split supersymmetry (SUSY). 
As a concrete example, we consider a SUSY $E_7/SU(5) \times U(1)^3$ non-linear sigma model in the framework of pure gravity mediation. The model accommodates three families of the quark and lepton chiral multiplets  as (pseudo) NG multiplets of the K{\" a}hler manifold, 
which may enable us to understand the origin and number of the families. 
We point out that squarks in the first and second generations are likely to be as light as a few TeV 
if the observed baryon asymmetry is explained by the thermal leptogenesis; 
therefore, these colored particles can be discovered at the LHC Run-2 or at the high luminosity LHC.
\end{abstract}

\clearpage

\section{Introduction}

It is well known that group theoretical properties and number of Nambu-Goldstone (NG) bosons are
determined by a given coset space $G/H$. If we extend it to a supersymmetric (SUSY) theory, the NG bosons are always accompanied by fermions. It is extremely interesting to identify the fermions (called as quasi NG fermions) with observed quarks and leptons \cite{Buchmuller:1982xn,Buchmuller:1982tf}, since it may provide us not only an origin of families of quarks and leptons but also an answer to the fundamental question  why we have three families in nature.

The SUSY $E_7$ non-linear sigma (NLS) model based on the coset space $E_7/SU(5) \times U(1)^3$~\cite{Kugo:1983ai, Yanagida:1985jc} is fascinating, since it can  accommodate three families of quarks and sleptons as its NG  chiral multiplets. This suggests that we may predict the maximal number of families in the approach of NLS based on exceptional groups, since the exceptional group is limited up to $E_8$. In fact, it is shown that
number of the families is limited to be three even if we take the biggest exceptional group $E_8$~\cite{Irie:1983cd}. 

In this paper we consider the $E_7/SU(5) \times U(1)^3$ model, where the unbroken $SU(5)$ is identified with 
the gauge group of the grand unified theory (GUT). 
The masses of the squarks and sleptons are highly suppressed at the tree level, 
since the global $E_7$ is assumed to be exact in the limit where gauge couplings and Yukawa couplings vanish.
Their masses are dominantly generated via radiative corrections, 
leading to a natural solution to the flavor changing neutral current (FCNC) problem. 
On the other hand, Higgs multiplets, $H_u$ and $H_d$, are considered to be non-NG multiplets, and hence, their masses are not suppressed.

The purpose of this paper is to examine if the above NG boson hypothesis of all the squarks and sleptons is consistent with observations. 
For this purpose, we choose one of the attractive and consistent mediation scenarios of SUSY breaking, that is called as pure gravity mediation (PGM)~\cite{Ibe:2006de,Ibe:2011aa} or minimal split SUSY~\cite{ArkaniHamed:2012gw}.
In the PGM,  gaugino masses are generated radiatively at the one-loop level via anomaly mediation~\cite{ Giudice:1998xp,Randall:1998uk}.
Since the gravitino mass is larger than $\mathcal{O}(100)$\,TeV, the cosmological gravitino problem is avoided easily. 
The gaugino masses are generated without a gauge singlet SUSY breaking (Polonyi) field; therefore, the cosmological Polonyi problem does not exist. 
Even without the gauge singlet SUSY breaking field, 
the Higgsino mass term of the order of the gravitino mass arises via an $R$-symmetry breaking term, i.e. the constant term in the superpotential~\cite{Inoue:1991rk, Ibe:2006de}. 
In the PGM, the lightest neutralino is the pure wino of a mass range of $\mathcal{O}(100$\,-$1000)$ GeV, providing us with a good dark matter candidate. Furthermore, it is expected that the SUSY FCNC problem is significantly relaxed.

In this paper, we point out that the $E_7/SU(5) \times U(1)^3$ NLS model is consistent with pure gravity mediation.\footnote{%
In Ref.~\cite{Harigaya:2015iva}, it has been shown that gaugino-Higgs mediation is consistent with the $E_7/SU(5) \times U(1)^3$ model. In this case, the higgsino-like neutralino rather than the stau can be the lightest SUSY particle, reducing the fine-tuning of the electroweak symmetry breaking. 
} 
The present model can be regarded as an ultraviolet completion of Higgs-Anomaly mediation proposed in Ref.~\cite{Yin:2016shg}. 
%In the $E_7$ NLS model together with pure gravity mediation, 
%the NG boson hypothesis for squarks and sleptons becomes convincing; 
%therefore, the prediction of the family number three based on the $E_7$ NLS model is plausible. 
Furthermore, we also show the predicted squark masses of the first and second generations are likely to be smaller than a few TeV when the thermal leptogenesis~\cite{Fukugita:1986hr} (see also \cite{Buchmuller:2005eh, Davidson:2008bu} for reviews)  successfully explains the observed baryon asymmetry. Those squarks can be discovered at the LHC Run-2 or at the high luminosity LHC. 

This paper is organized as follows. In Sec.\,2, we explain our setup, the $E_7$ NLS model combined with PGM. 
It is shown that the NG multiplets of the three chiral generations have vanishing masses at the tree level.
In Sec.\,3, we investigate low-energy phenomenology of the model, paying attention to the the lightest squark and slepton masses. Section 4 is devoted to the conclusion.

\section{$E_7/SU(5) \times U(1)^3$ model in pure gravity mediation}

The $E_7/SU(5) \times U(1)^3$ model is obtained via $E_7/SU(5) \times SU(3) \times U(1)$ model, 
which contains $133-24-8-1=100$ NG modes. The NG multiplets are~\cite{Kugo:1983ai} 
\begin{eqnarray}
\phi_a^i: ({\bf \bar 5, 3}, 2), \ 
\phi_i^{ab}: ({\bf 10, \bar 3}, 1), \ 
\phi^a: ({\bf 5,1},3),
\end{eqnarray}
where $a,b=1 \dots 5$ and $i=1 \dots 3$. Note that $\phi_a^i$ and $\phi_i^{ab}$ are identified with chiral multiplets of three family leptons and quarks, whose scalar components are massless at the tree level.
Here, we assume $\phi^a$ has a large Dirac mass term with another ${\bf \bar 5'}$,\footnote{
We do not identify the NG multiplet  $\phi^a$  with one of Higgs multiplets, $H_u$, in this paper.
}
 which is required to cancel the non-linear sigma model and gauge anomalies~\cite{Yanagida:1985jc}.
The Higgs chiral multiplets, $H_u$ and $H_d$, are introduced as non-NG matter multiplets. 
Then, their boson components are expected to have SUSY breaking soft masses of the order of the gravitino mass, $m_{3/2}$.
The $E_7/SU(5) \times U(1)^3$ model is obtained through the further breaking $SU(3)$ down to $U(1)^2$, leading to six more NG bosons. 
These NG bosons are identified with scalar partners of three right-handed neutrinos. 
After all, the three new NG chiral multiplets of $E_7/SU(5)\times U(1)^3$ are identified with three chiral multiplets of the right-handed neutrinos.

The K{\" a}hler potential for the NG multiplets is constructed from a real function transforming under $E_7$ as
\begin{eqnarray}
\mathcal{K}(\phi, \phi^\dag) \to \mathcal{K}(\phi, \phi^\dag) + f_H(\phi) + f_H(\phi)^\dag,
\end{eqnarray}
where $f_H$ is a holomorphic function of $\phi$, and $\mathcal{K}$ is invariant under transformations of the unbroken symmetry, $SU(5) \times U(1)^3$. The real function $\mathcal{K}$ itself can not be $E_7$ invariant, since the shift of $\mathcal{K}$ with the holomorphic function does not leave the Lagrangian invariant in supergravity theories: we need an chiral superfield, $S$, to cancel the shift~\cite{Komargodski:2010rb, Kugo:2010fs}. Then, we have the $E_7$ invariant K{\" a}hler potential as
\begin{eqnarray}
K(\phi,\phi^\dag,S,S^\dag) = F(\mathcal{K}(\phi, \phi^\dag)  + S + S^\dag),
\end{eqnarray}
with
\begin{eqnarray}
S \to S - f_H(\phi),
\end{eqnarray}
under $E_7$ transformation.

\paragraph{Soft masses}
To examine a soft SUSY breaking mass for $\phi$, let us consider the leading term of $\phi^\dag \phi$ as
\begin{eqnarray}
K = F( \phi^\dag \phi + S + S^\dag + \dots),
\end{eqnarray}
where $\dots$ denotes irrelevant terms such as higher order terms of $\phi^\dag \phi$. 

The soft SUSY breaking mass of $\phi$ arises from
\begin{eqnarray}
V(S,S^\dag, \phi, \phi^\dag)  &=& e^{K/M_P^2} \left[ \left|\frac{\partial K}{\partial \phi}\right|^2 K^{-1}_{\phi\phi} +
\left|\frac{\partial K}{\partial S}\right|^2 K^{-1}_{SS} + 
\frac{\partial K}{\partial \phi}\frac{\partial K}{\partial S^\dag} K^{-1}_{\phi S} + h.c. \right] \frac{|W|^2}{M_P^4} \nonumber \\ 
&=& e^{K/M_P^2}  \left(\frac{\partial K}{\partial x}\right)^2 \left(\frac{\partial^2 K}{\partial x^2}\right)^{-1} \frac{|W|^2}{M_P^4}
= G(x) , \label{eq:gs_mass}
\end{eqnarray}
where $x=\phi^\dag \phi + S + S^\dag$; $W$ is the superpotential with $e^{K/M_P^2} |W|^2 = m_{3/2}^2$. 
Here, $M_P \simeq 2.4 \times 10^{18}$\,GeV is the reduced Planck mass.
Then,
\begin{eqnarray}
\frac{\partial^2 V}{\partial \phi^\dag \partial \phi} = \frac{\partial V}{\partial x} + |\phi|^2 \frac{\partial^2 V}{\partial x^2},
\end{eqnarray}
which vanishes at the minimum. We see all NG bosons are massless at the tree level.\footnote{
See also Ref.~\cite{goto_yanagida} for another clarification of the masslessness.
} The above argument does not depend on whether there is a direct coupling between a (pseudo) NG multiplet and a SUSY breaking field $Z$,
since the potential is just replaced as $G(x) \to G(x, Z, Z^\dag)$ (see appendix for an explicit calculation).

On the other hand, we assume that the soft SUSY breaking masses for the Higgs doublets, which are non-NG multiplets, 
are given by:
\begin{eqnarray}
m_{H_u}^2 = m_{H_d}^2 \simeq c_H m_{3/2}^2, \label{eq:higgs_soft}
\end{eqnarray}
at the tree level, where $c_H$ is a constant free parameter. The negativeness of $m_{H_u}^2$ and $m_{H_d}^2$ is 
important to solve the tachyonic slepton problem~\cite{Yin:2016shg} and give large masses for stops~\cite{Yamaguchi:2016oqz, Yin:2016shg}, enhancing the 
Higgs boson mass~\cite{Okada:1990vk,  Ellis:1990nz,  Haber:1990aw, Okada:1990gg, Ellis:1991zd}. 
%$c_H<0$ does not lead to a tachyonic vacuum, if the Higgs mixing term, $\mu$, is large enough.
%
In fact, the soft SUSY breaking masses for NG bosons arise radiatively via anomaly mediation 
and renormalization group (RG) running effects from $m_{H_u}^2$ and $m_{H_d}^2$. 
At the high energy scale, say, the GUT scale ($M_{\rm GUT}$), the scalar mass of the NG multiplet is written as
\begin{eqnarray}
\tilde m_{\phi}^2 (\mu_R = M_{\rm GUT}) \simeq - \frac{1}{4} \Bigl[ 
\frac{\partial \gamma_\phi}{\partial g} \beta_g + \frac{\partial \gamma_{\phi}}{\partial y} \beta_y 
\Bigr] m_{3/2}^2 \, , \label{eq:amsb_scalar}
\end{eqnarray}
where $m_{3/2}$ is a gravitino mass and $\gamma_\phi$ is an anomalous dimension defined by $\gamma_\phi \equiv {\partial \ln Z_{\phi}}/{\partial \ln \mu_R}$ with $\mu_R$ of a renormalization scale. For $c_H=0$ in Eq.\,(\ref{eq:higgs_soft}), 
the above relation holds at any scale and the sleptons are tachyonic below the GUT scale due to the positive beta-functions for the $U(1)_Y$ and $SU(2)_L$ gauge couplings. In the next section, we show that this problem can be solved if the Higgs soft mass squares are  
negative at $M_{\rm GUT}$~\cite{Yin:2016shg}. 

Since there is no Polonyi field in our setup, the gaugino masses are purely generated from anomaly mediation:
\begin{eqnarray}
M_{1} = \frac{33}{5} g_1^2 m_{3/2}, \ M_{2} = g_2^2 m_{3/2}, \  M_{3} = -3 g_3^2 m_{3/2},
\end{eqnarray}
above a SUSY particle mass scale. Here, $M_{1}, M_2$ and $M_3$ are bino, wino and gluino mass, respectively; 
$g_1$, $g_2$ and $g_3$ are gauge coupling constants of $U(1)_Y$, $SU(2)_L$ and $SU(3)_C$.
Below the SUSY particle mass scale, the threshold corrections~\cite{Pierce:1996zz} need to be included. Trilinear couplings are also generated from anomaly mediation and given by
\begin{eqnarray}
A_{l m n}  = \frac{1}{2} (\gamma_l + \gamma_m + \gamma_n)  m_{3/2} ,
\end{eqnarray}
with $V \ni A_{l m n} y_{lmn} \phi_l \phi_m \phi_n + h.c.$

\section{Phenomenological implications}

\subsection{Case of vanishing squark and slepton masses} \label{sec:vanish}
We assume, in this subsection, that the gauge and Yukawa interactions are only sources of the explicit breaking of the global $E_7$, and hence all squarks and sleptons are massless at the tree level.\footnote{
The quantum corrections other than those from anomaly mediation may be suppressed enough, 
if one consider the sequestering between the SUSY breaking hidden sector and the MSSM sector (see e.g. appendix in Ref.~\cite{hyy_gaugino}).  
} 
The masses of squark and sleptons arise only through anomaly mediation at the high energy (GUT) scale.
 Thus, sleptons are tachyonic at $M_{\rm GUT}$ as discussed in the previous section.  However, radiative corrections from the Higgs loops lift up the slepton mass for $c_H<0$, solving the tachyonic slepton problem. We estimate the positive contributions to sfermion masses from Higgs loops by taking into account RG running effects. The resultant mass spectrum is hierarchical among generations since Yukawa couplings of Higgs doublets are hierarchical~\cite{Yamaguchi:2016oqz,Yin:2016shg}, and hence we obtain relatively large masses for stops
which explains easily the observed Higgs boson mass of $125$\,GeV.

Since the sleptons and squarks are massless at the tree level, this model has four parameters:
\begin{eqnarray}
m_{3/2}, \  \tan\beta   ,\ c_H, \ {\rm sign}(\mu),
\end{eqnarray}
where $\tan\beta$ is the ratio of the vacuum expectation values of $H_u$ to $H_d$. Note that from the conditions of the correct electroweak symmetry breaking, 
the viable range of $\tan\beta$ is fixed to be $\tan\beta \gtrsim 40$ (60) for ${\rm sign}(\mu) = +\, (-)$.
This is because one needs large Yukawa couplings of the tau and bottom, $Y_{\tau}$ and $Y_b$.

The necessity of the large $\tan\beta$ is explained as follows. From the electroweak symmetry breaking (EWSB) conditions, we have
\begin{eqnarray}
\label{eq:A_Higgs}
m_A^2 \simeq (m_{H_u}^2 + m_{H_d}^2 + 2 |\mu|^2) \simeq  (m_{H_d}^2 - m_{H_u}^2 )(1 + 2/\tan^2\beta),
\end{eqnarray}
where $m_A$ is a mass of the CP-odd Higgs.  
Here, we have used $\mu^2 \simeq -m_{H_u}^2 + (m_{H_d}^2 - m_{H_u}^2)/\tan^2\beta$.
In order to avoid a tachyonic CP-odd Higgs, $m_A^2<0$, 
$m_{H_d}^2 > m_{H_u}^2$ should be satisfied at the electroweak scale. 
This is achieved via the RG running of $m_{H_d}^2$,
\begin{eqnarray}
\frac{d m_{H_d}^2}{d \ln \mu_R} \ni \frac{1}{16\pi^2} (6 Y_b^2 + 2 Y_{\tau}^2 )m_{H_d}^2,
\end{eqnarray}
for large $Y_b$ and $Y_{\tau}$ (remember that $m_{H_d}^2$ is negative). For large $\tan\beta \gtrsim 40$ (or 60 for $\mu<0$), the above contribution is larger than that for $m_{H_u}^2$,
\begin{eqnarray}
\frac{d m_{H_u}^2}{d \ln \mu_R} \ni \frac{1}{16\pi^2} (6 Y_t^2 )m_{H_u}^2.
\end{eqnarray}
Then, $m_{H_d}^2 > m_{H_u}^2$ ($|m_{H_d}^2| < |m_{H_u}^2|$) is realized at the low-energy scale, which is required for the correct EWSB.

%%%%%%%%%%%%
\begin{figure}[!t]
\begin{center}
\includegraphics[bb=100 150 600 650, scale=0.4]{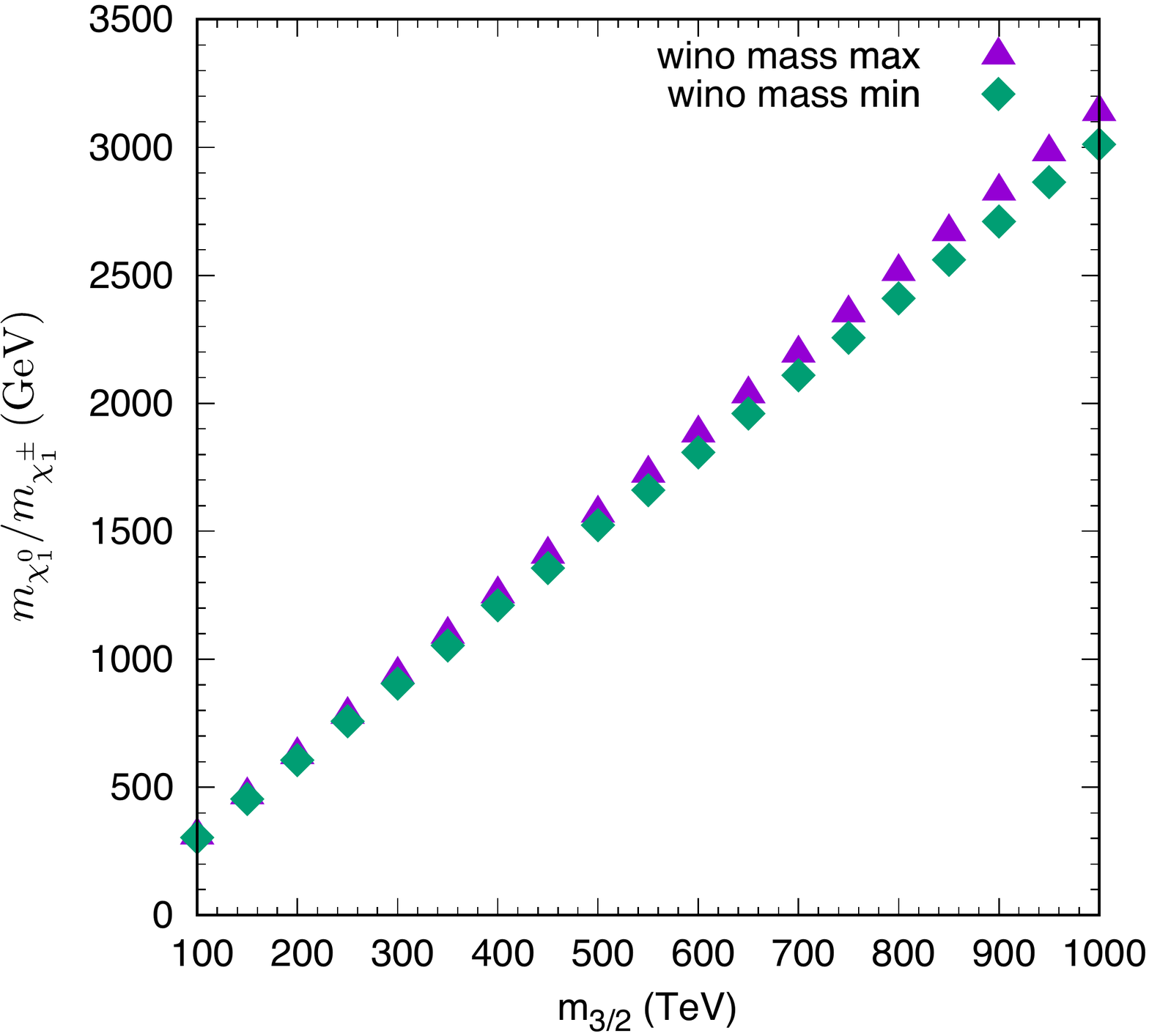}
\includegraphics[bb=0 150 500 650, scale=0.4]{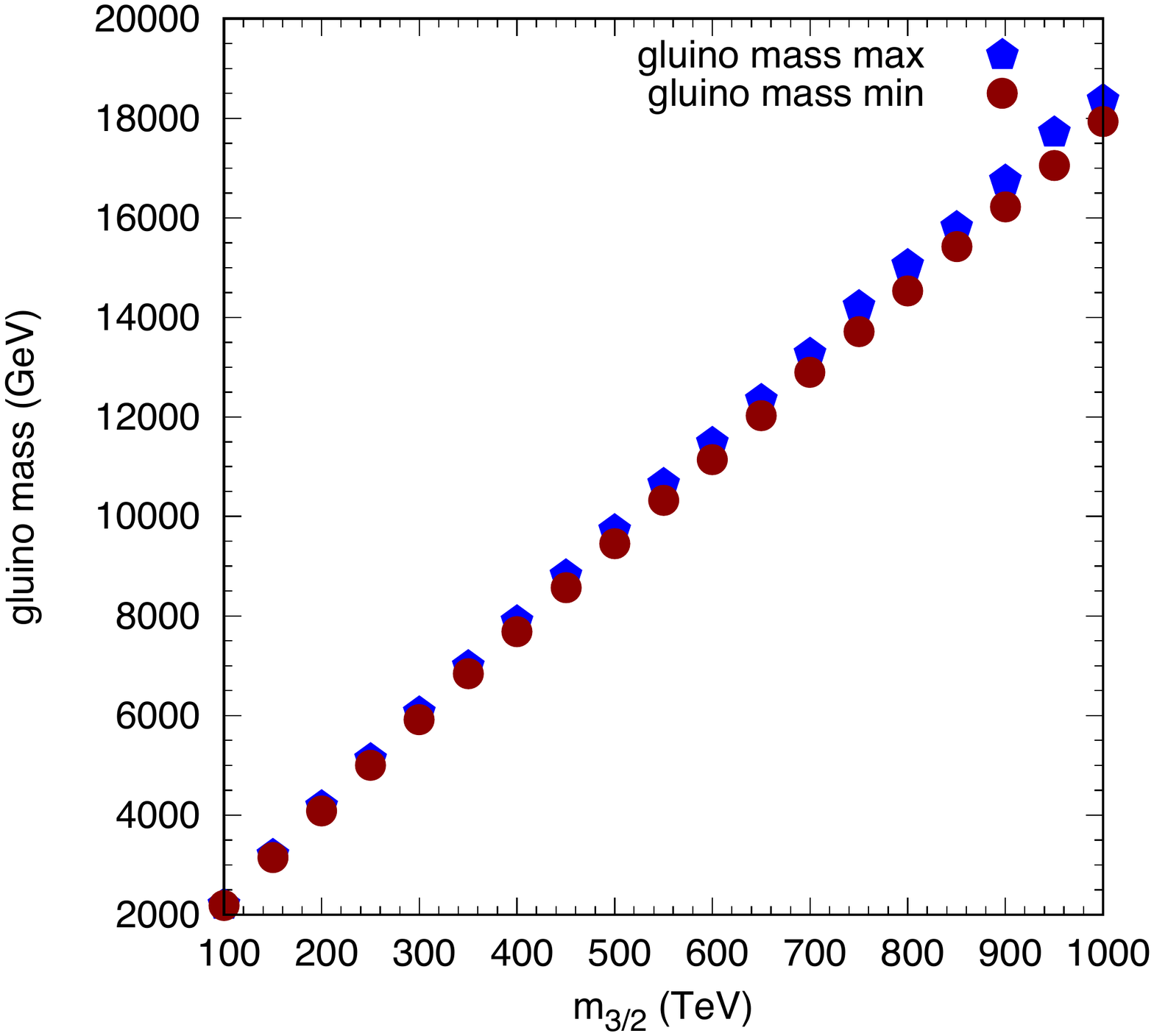}
\caption{
The masses of the lightest chargino/neutralino and the gluino in the case of the vanishing squark and slepton masses. 
The triangles and pentagons (squares and circles) show maximum (minimum) values. 
Here, $\alpha_s(m_Z)=0.1185$ and $m_t({\rm pole})=173.34$\,GeV.
%In the red (green) region, the muon $g-2$ is explained at 1$\sigma$ (2$\sigma$) level.
}
\label{fig:1}
\end{center}
\end{figure}
%%%%%%%%%

The dependence of the required $\tan\beta$ on ${\rm sign}(\mu)$ originates from the radiative correction to $Y_b$ from sbottom-gluino loops~\cite{Hall:1993gn,Carena:1994bv}. 
Since the SUSY contribution to the muon $g-2$ is opposite to the needed one and the required $\tan\beta$ for correct EWSB is so large in the case of ${\rm sign}(\mu)=-$, we focus on the region of ${\rm sign}(\mu) = +$ in this paper, unless otherwise noted. In fact, we can not find particular differences in the SUSY-particle spectra between the regions with ${\rm sign}(\mu) = -$ and sign$(\mu)=+$, apart from the required $\tan\beta$ and the contribution to the muon $g-2$.

 The SUSY spectrum is computed using {\tt SuSpect\,2.4.3}~\cite{Djouadi:2002ze}. We scan all allowed parameter space of $c_H$ and $\tan\beta$, 
\begin{eqnarray}
0 >  c_H > -0.5, \ \ 40< \tan\beta < 60, \label{eq:scan_pars}
\end{eqnarray}
 and take the maximum and minimum values of $m_{\chi_1}^\pm/m_{\chi_1^0}$ for each $m_{3/2}$. Here, $m_{\chi_1}^\pm$ and $m_{\chi_1^0}$ are the lightest chargino mass and neutralino mass, respectively, and they are nearly degenerated. The input parameter $c_H$ is set at $M_{\rm inp}=10^{16}$\,GeV ($\approx M_{\rm GUT}$).
We have demanded the lightest slepton (squark) to be heavier than 340 (1000) GeV from the null results of the LHC SUSY searches~\cite{Chatrchyan:2013oca, ATLAS_jets}. \footnote{
The lower bound of the slepton mass of 340\,{\rm GeV} is given in the case that the slepton is lighter than the lightest neutralino and effectively stable inside the detectors.}

The range of $c_H$ has been determined to avoid the tachyonic squarks and sleptons: squarks become tachyonic in the regions of $c_H<-0.5$.  Note that on the viable parameter space, we find that all Yukawa couplings at the GUT scale are smaller than $\sqrt{0.1 \times 4\pi}$. 

We now show the mass of the lightest chargino, which is almost pure wino, in Fig.~\ref{fig:1} (left). 
We see the wino mass is almost insensitive to $\tan\beta$ and $c_H$. 
This is because the threshold correction~\cite{Pierce:1996zz}
\begin{eqnarray}
\frac{\Delta M_2}{M_2} \ni \frac{g_2^2}{16\pi^2} \frac{\mu}{M_2} \sin(2\beta) \frac{m_A^2}{\mu^2-m_A^2}\ln\frac{\mu^2}{m_A^2}
\end{eqnarray}
is suppressed by large $\tan\beta$ and $\mu^2 \gg m_A^2$: in contrast to usual PGM models, the relation between wino mass and gravitino mass is almost fixed.
Since the Higgsino is heavy, the leading dark matter candidate in this model is the wino-like neutralino. 
This is the reason why we restrict the range of $m_{3/2} \simeq$\,100\,-1000\,TeV, 
(approximately) corresponding to the mass range for the wino dark matter, 270\,GeV\,$< m_{\tilde \chi_1^0} <$\,2.9 TeV.
This is followed by the phenomenological constraints assuming that the wino is the lightest SUSY particle (LSP). 
The lower-bound comes from the wino search at the LHC~\cite{Aad:2013yna} 
while the upper-bound is given by the condition to avoid the over-closure of the universe with the wino dark matter~\cite{Bhattacherjee:2014dya}.\footnote{
In our case only some of the squarks are light as 1000 GeV, i.e. the event cross sections are expected to be smaller than those used in the reference. 
%
%We demand the slepton be heavier than $m_{\chi_1^0}$ in a small region where $m_{\chi_1^0} < 340$ GeV.
}

On the right panel in Fig.~\ref{fig:1}, we also show the maximum and minimum values of the gluino mass. 
The gluino is accessible at the LHC only when the gravitino mass is smaller than about 150\,TeV.

%%%%%%%%%%%%
\begin{figure}[!t]
\begin{center}
\includegraphics[scale=1.2]{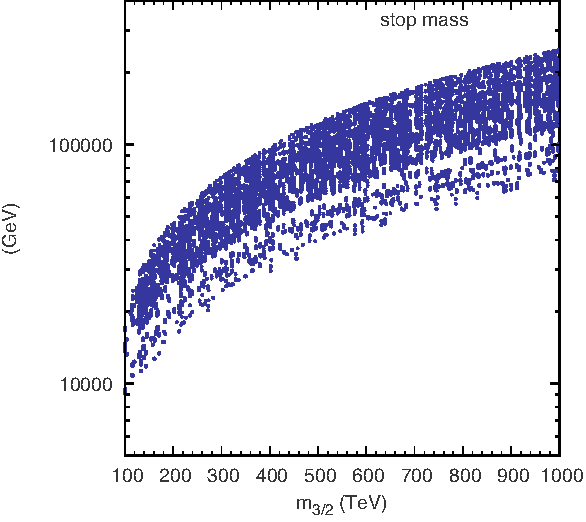}
\caption{
Scatter plot of the stop mass.
}
\label{fig:mstop}
\end{center}
\end{figure}
%%%%%%%%%

%%%%%%%%%%%%
\begin{figure}[!t]
\begin{center}
\includegraphics[scale=1.05]{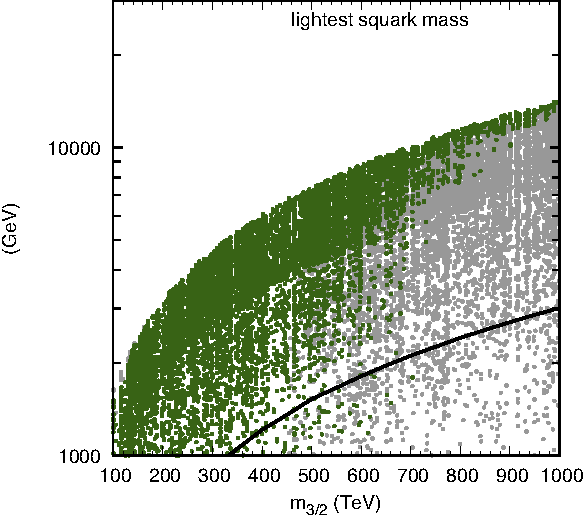}
\includegraphics[scale=1.05]{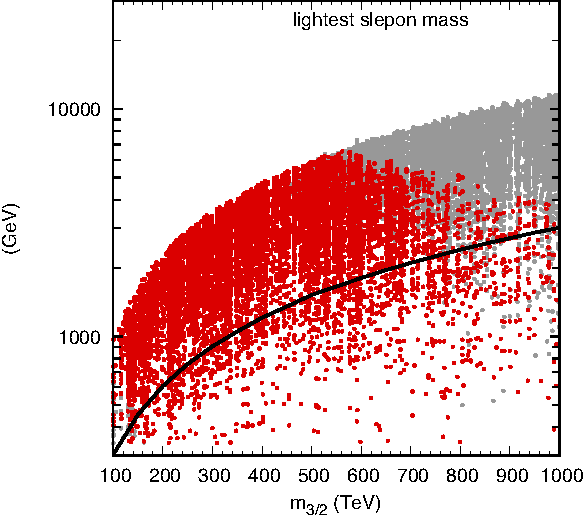}
\caption{
Scatter plots of the lightest squark mass (left) and the lightest slepton mass (right).
The dark-green and red dots satisfy the constraint $122\,{\rm GeV} \leq m_h \leq 128\,{\rm GeV}$ while the gray dots do not. The black solid lines show the minimum values of $m_{\chi_1^0}$
%In the red (green) region, the muon $g-2$ is explained at 1$\sigma$ (2$\sigma$) level.
}
\label{fig:2}
\end{center}
\end{figure}
%%%%%%%%%

%%%%%%%%%%%%
\begin{figure}[!t]
\begin{center}
\includegraphics[scale=0.63]{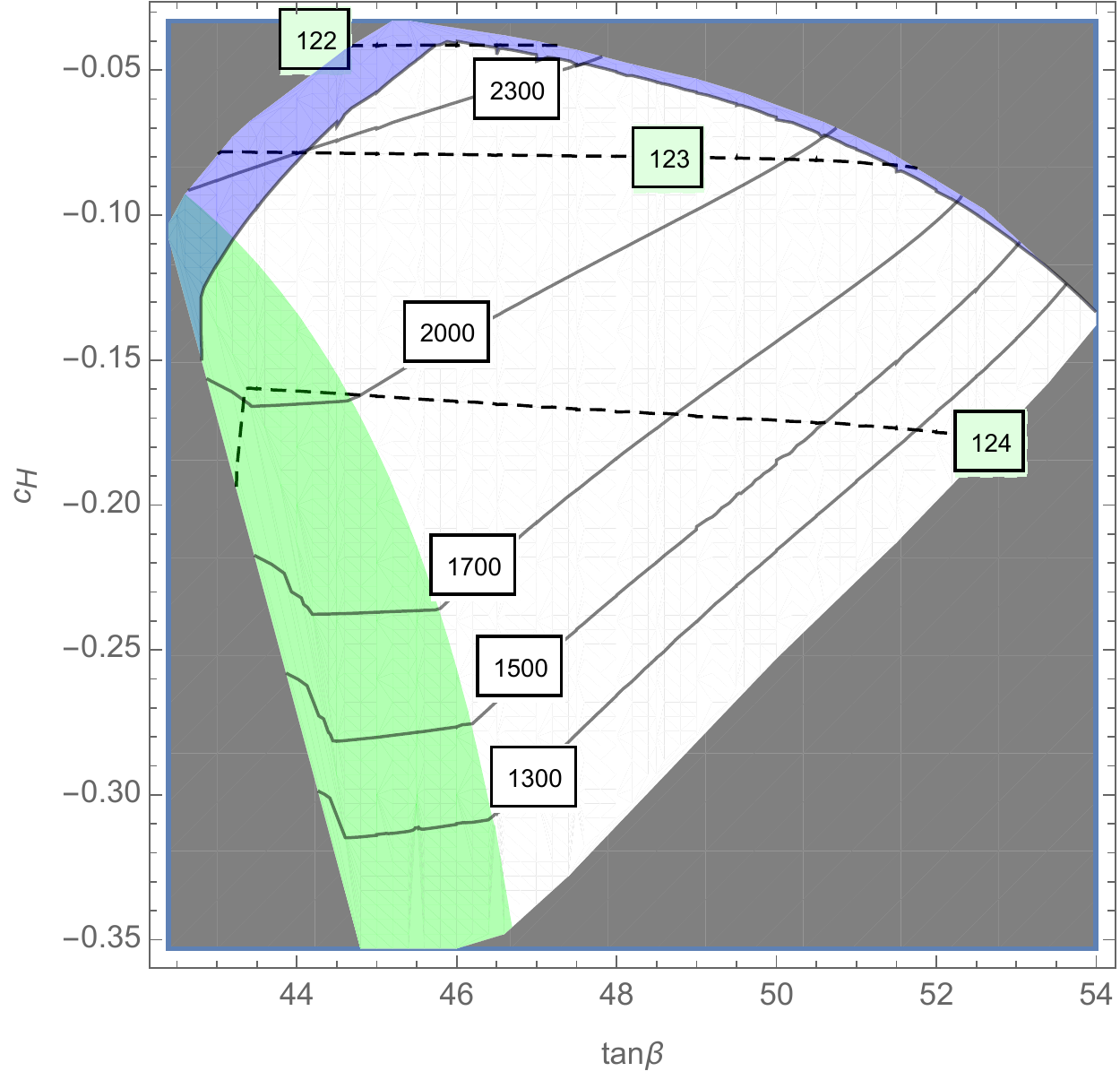}
\includegraphics[scale=0.63]{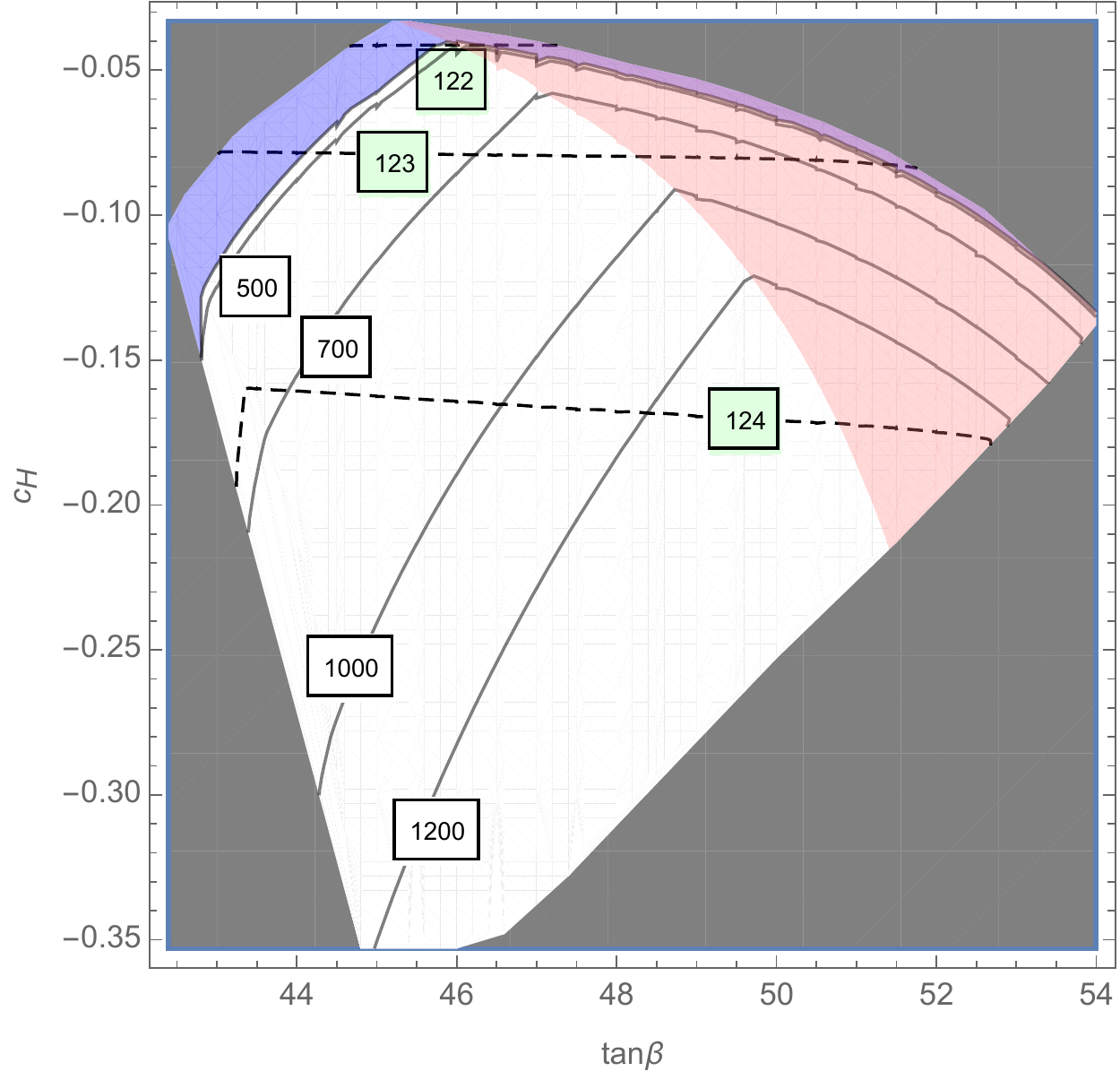}
\includegraphics[scale=0.63]{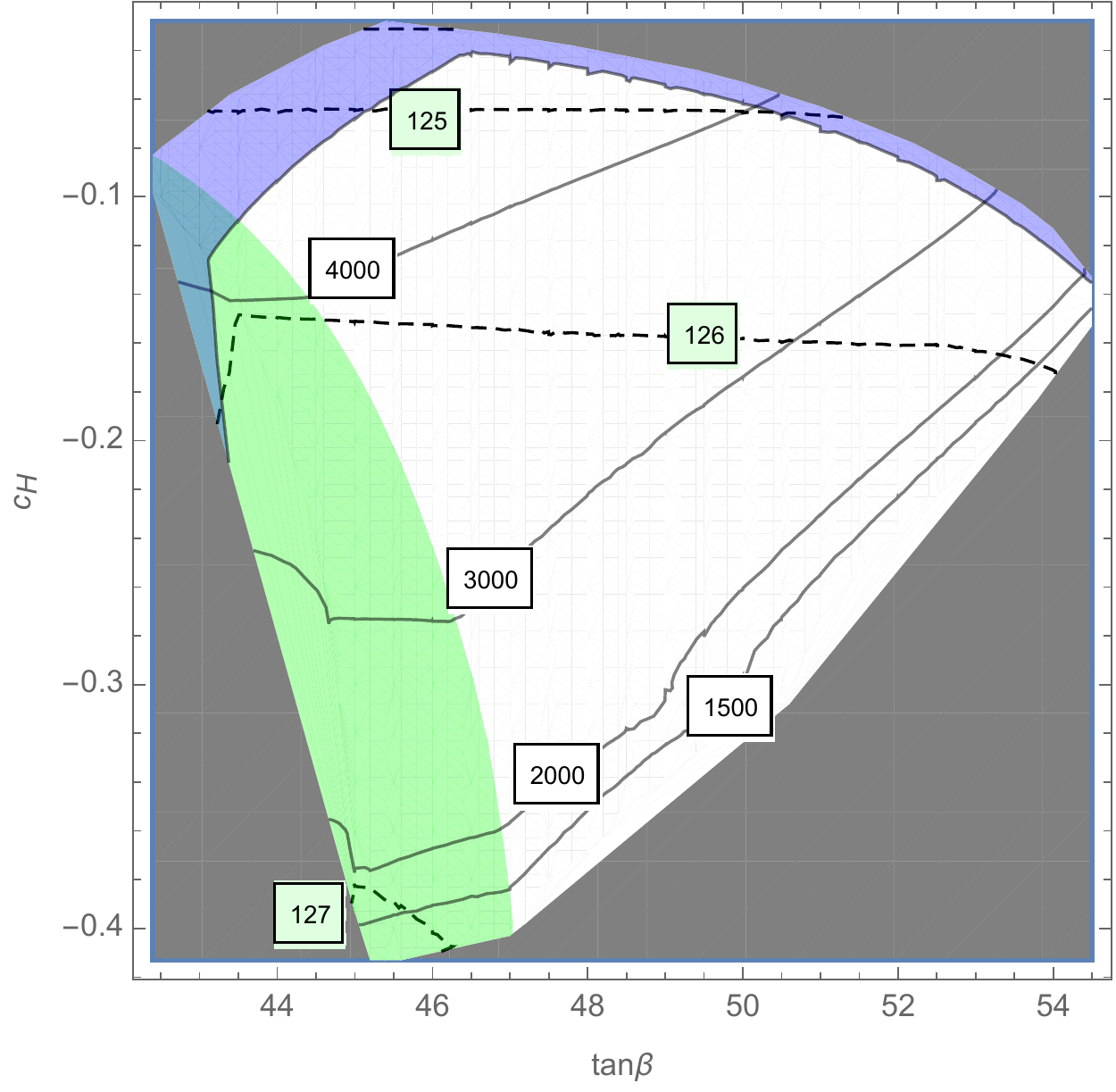}
\includegraphics[scale=0.63]{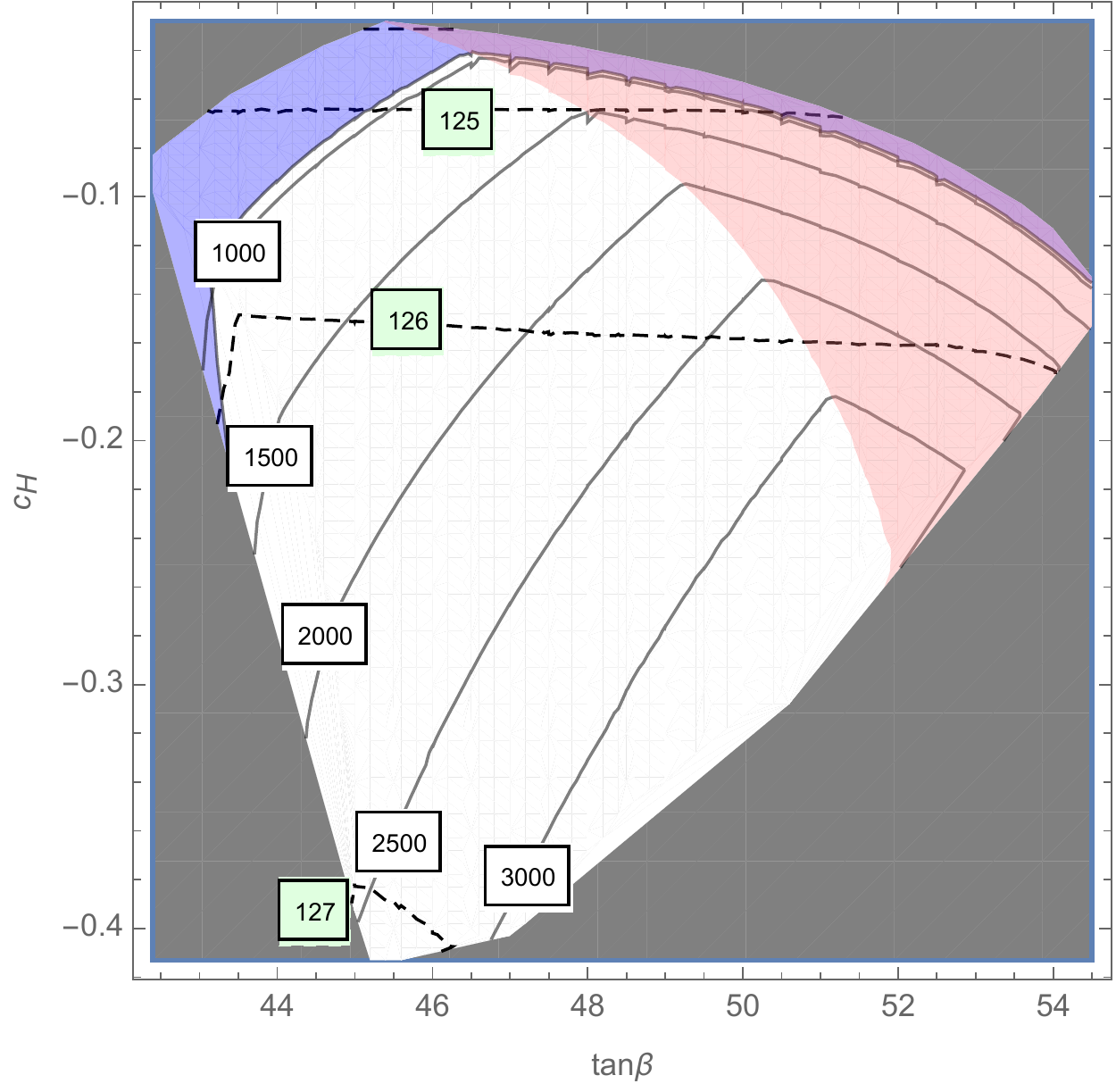}
\caption{
Contours of the lightest squark mass (left panels) and slepton mass (right panels)  in unit of GeV.
The upper (lower) two panels the gravitino mass is taken as $m_{3/2}=150$ (300)\,TeV. 
The Higgs boson masses in unit of GeV are shown as dashed lines. 
In the blue shaded region, the selectron is the LSP.
In the green (pink) shaded regions, 
the lightest squark (slepton) is the right-handed down squark (the left-handed selectron).
}
\label{fig:cont1}
\end{center}
\end{figure}
%%%%%%%%%

%%%%%%%%%%%%
\begin{figure}[!t]
\begin{center}
\includegraphics[scale=0.63]{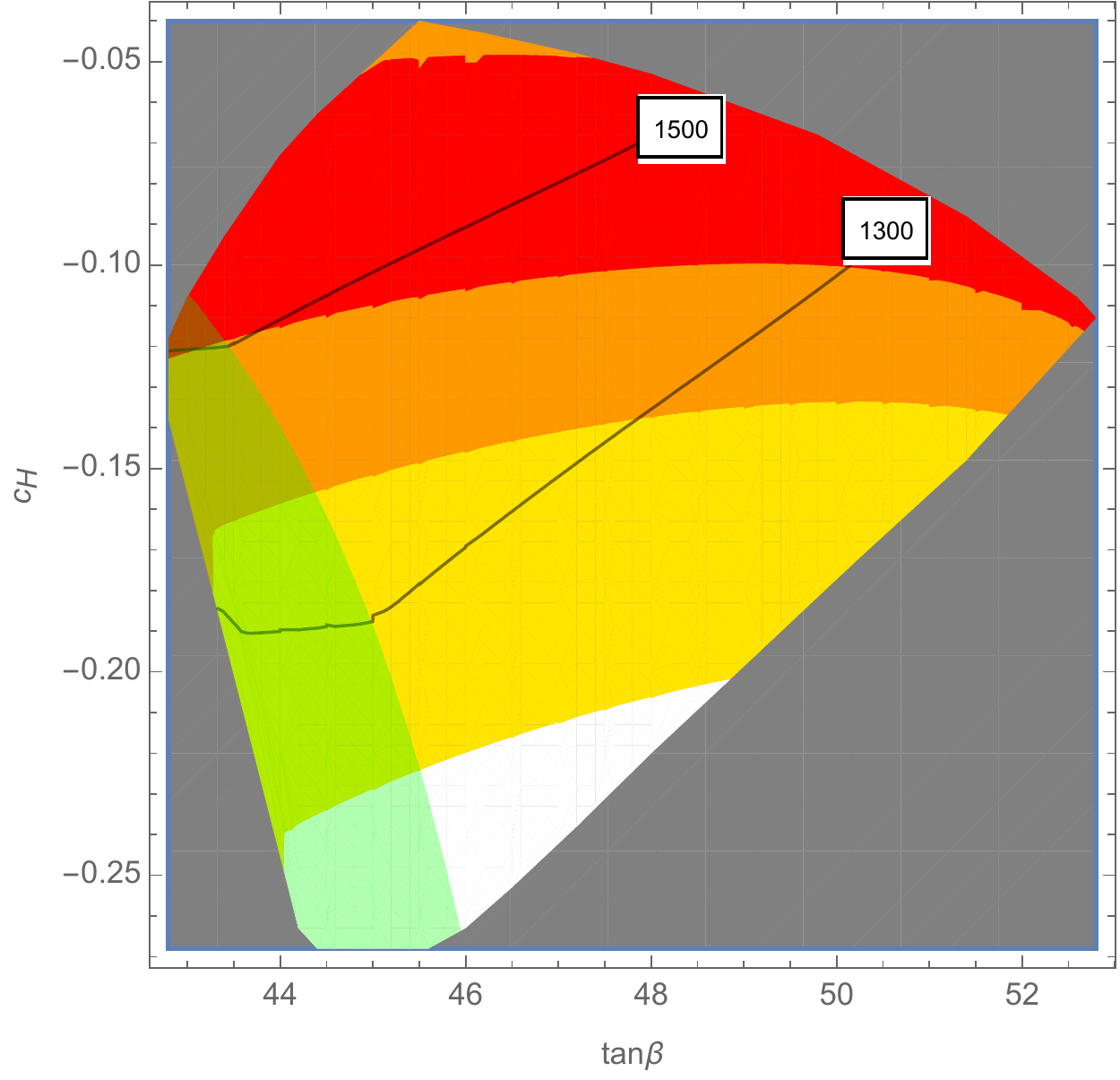}
\includegraphics[scale=0.63]{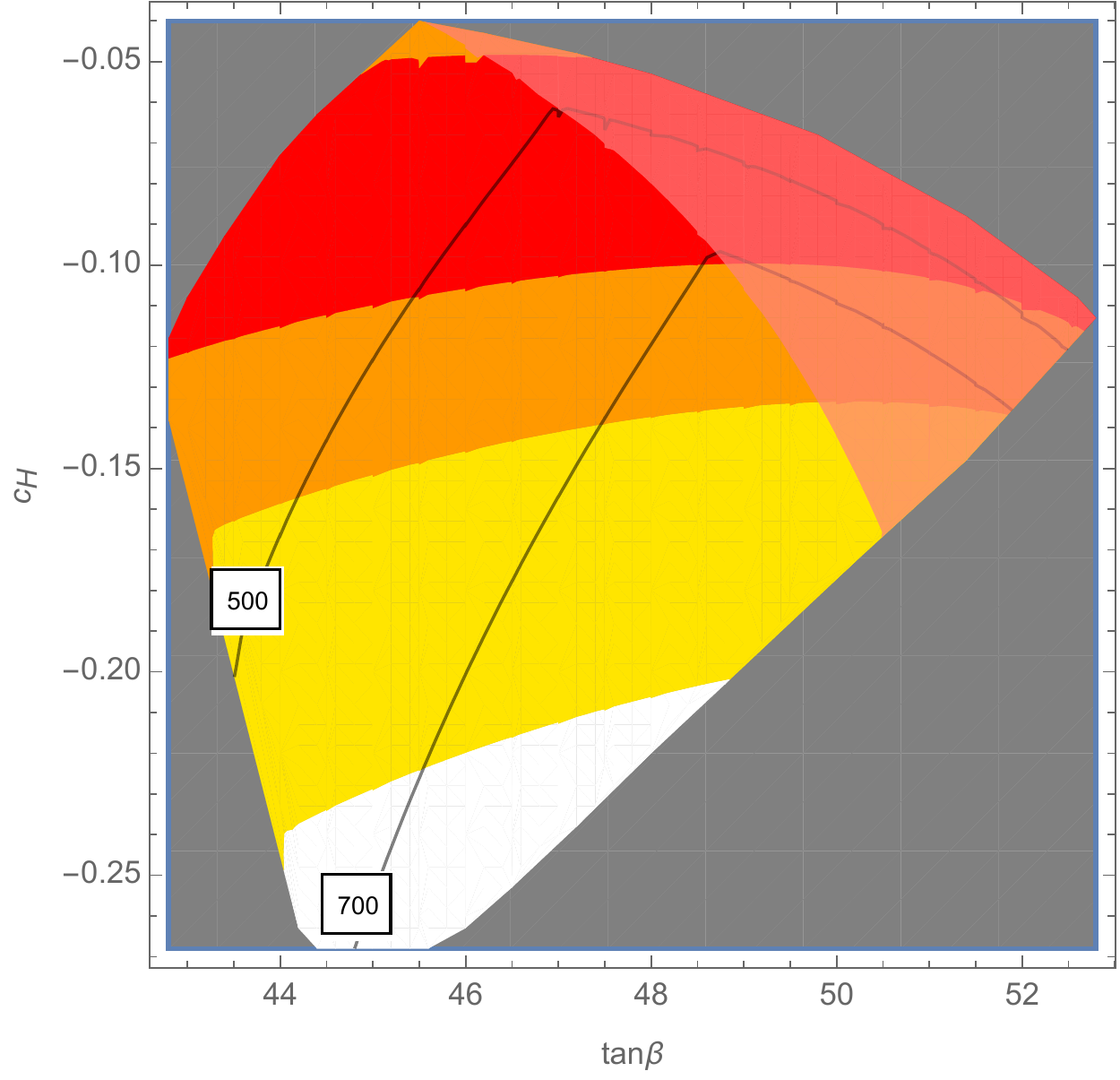}
\includegraphics[scale=0.63]{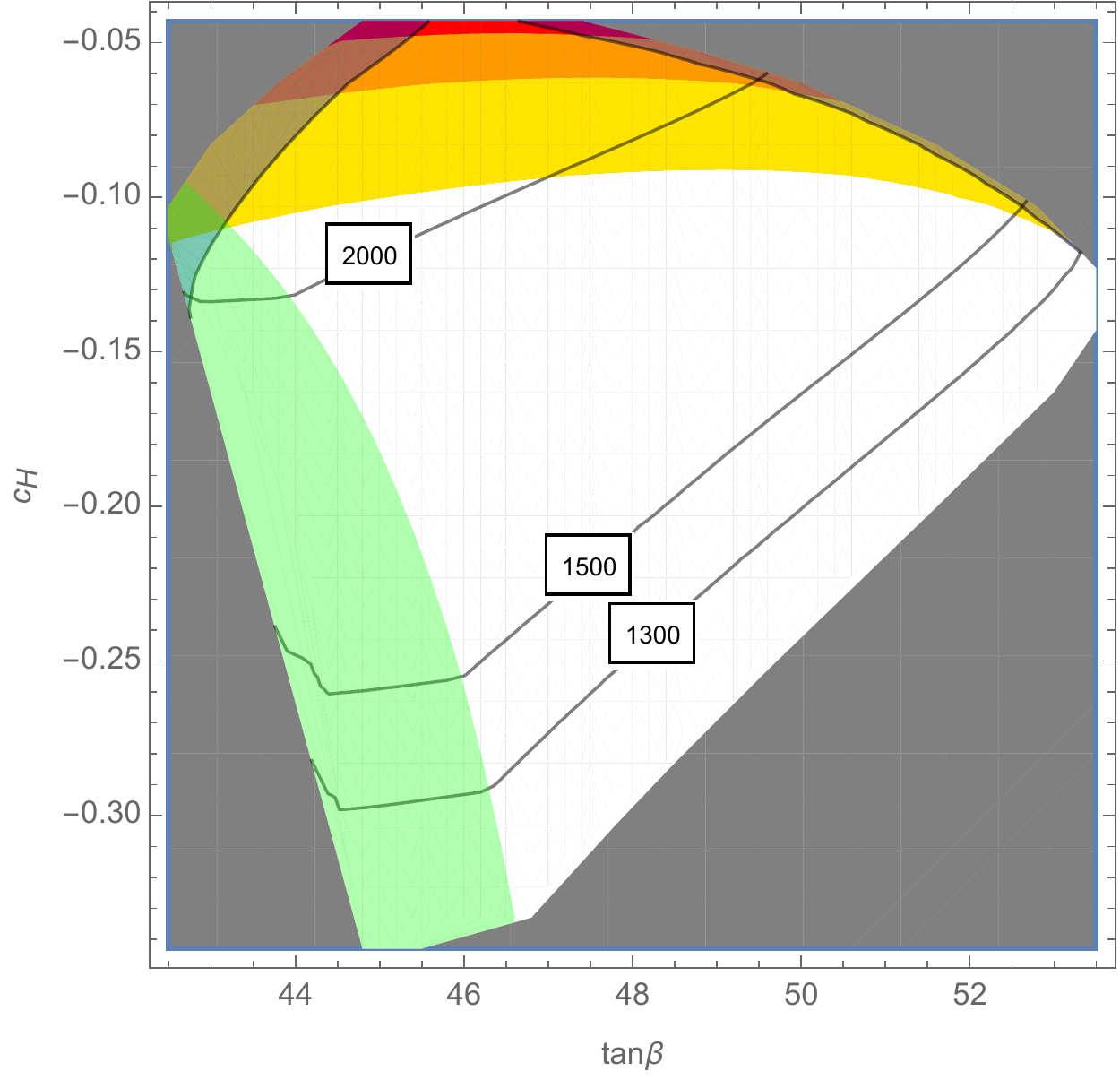}
\includegraphics[scale=0.63]{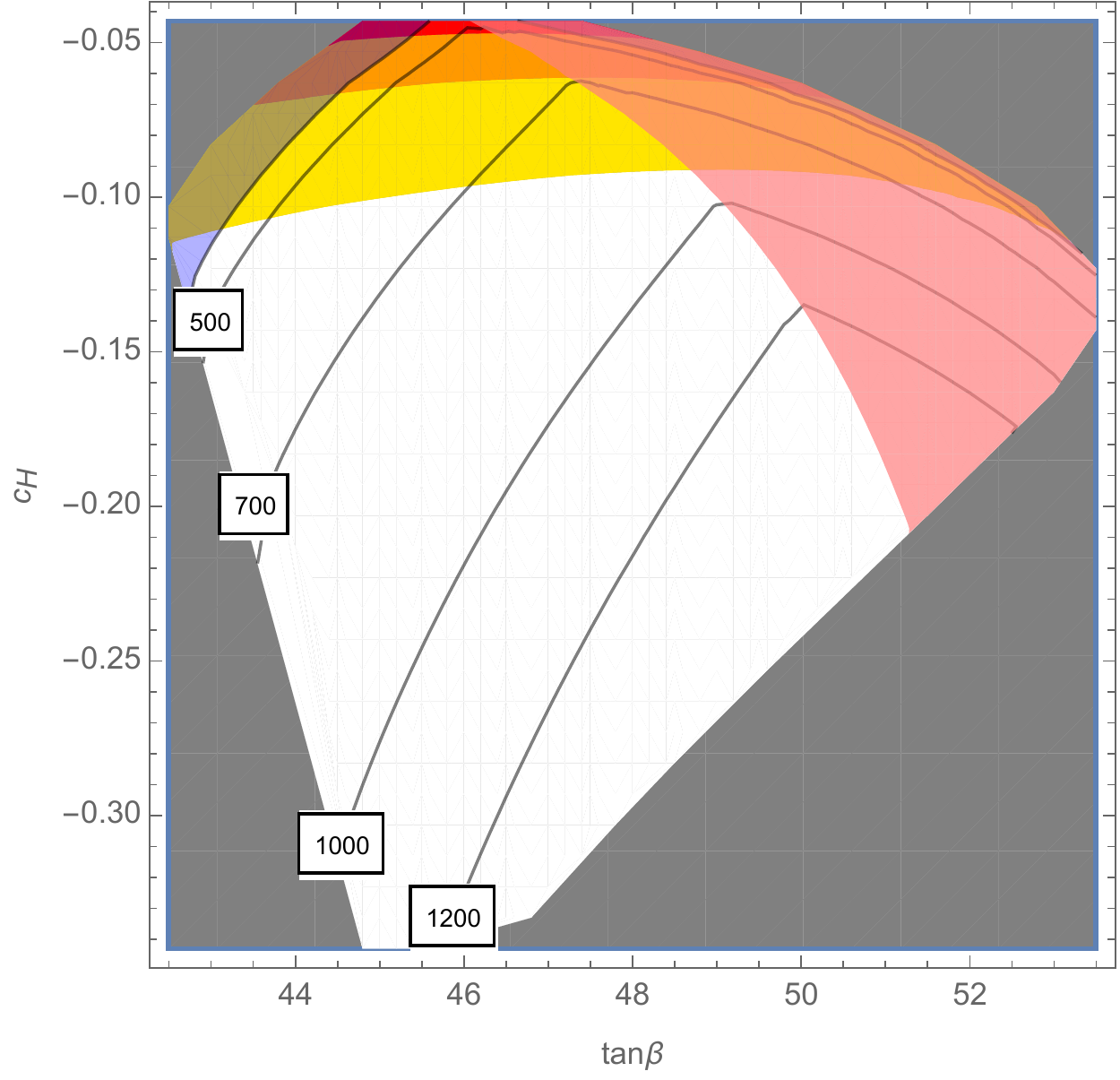}
\caption{
The regions consistent with the muon $g-2$ experiment at $1\sigma$ (red), 1.5$\sigma$ (orange) and 2$\sigma$ (yellow) level. 
In the upper two panels, the gravitino mass is taken as $m_{3/2}=100$ (140) TeV. 
We also show the contours of the lightest squark (slepton) mass in the left (right) panels, in unit of GeV. 
In the blue shaded regions, the selectron is the LSP.
}
\label{fig:g-2}
\end{center}
\end{figure}
%%%%%%%%%

Next, we show the stop mass, defined by $m_{\tilde t} \equiv (m_{Q_3} m_{\bar U_3})^{1/2}$ in Fig.~\ref{fig:mstop}
to estimate regions consistent with the Higgs boson mass of 125 GeV. 
Here, $m_{Q_3}$ and $m_{\bar U_3}$ are the masses of the left-handed and right-handed stop, respectively.  
The scanned range of the parameter space is the same as in Eq.\,(\ref{eq:scan_pars}).
The stop mass is lifted up via RG running:
\begin{eqnarray}
\frac{d m_{Q_3^2}}{d\ln \mu_R} \ni \frac{1}{16\pi^2} 2 Y_t^2 m_{H_u}^2, \ \ 
\frac{d m_{\bar U_3}^2}{d\ln \mu_R} \ni \frac{1}{16\pi^2} 4 Y_t^2 m_{H_u}^2 \,.
\end{eqnarray}
Therefore, the large $|c_H|$ leads to the larger stop masses. 
On almost all points, the stop mass is larger than 10\,TeV,~\footnote{
Only in the case that $|c_H|$ is as small as 0.04 and $m_{3/2}=$100\,TeV, the stop mass becomes $\sim 9$\,TeV.
} and hence the Higgs boson mass of 125\,GeV is easily explained in most of the parameter space.

In Fig.~\ref{fig:2}, the scatter plots of the masses of the lightest squark and slepton are shown. 
Here, the lightest squark is either the right-handed up squark or down squark,
%\footnote{
%
%The right-handed strange squark is also light in this case.
%
%} 
while the lightest slepton is the left-handed or right-handed selectron, depending on the parameter space.
The threshold corrections to the squark masses are included utilizing one-loop RG equations (see e.g. \cite{Box:2008xu}) to incorporate the large mass hierarchy between the third generation squarks and first/second generation squarks.\footnote{
We have modified the {\tt SuSpect} code to include the resummation of the logarithmic corrections.
}
The dark-green and red dots satisfy the constraint, $122 \le m_h \le 128$\,GeV, while the gray dots do not. Here, $m_h$ is the Higgs boson mass computed using {\tt SUSYHD 1.0.2}~\cite{Vega:2015fna}. The black solid lines show the minimum values of $m_{\chi_1^0}$ in Fig. \ref{fig:1}. Thus, the points below the lines are excluded unless the $R$-parity is violated.
It can be seen that the lightest squark can be as light as 1\,-\,2\,TeV even for $m_{3/2}=1000$\,TeV. 
However, the stop mass is too large in such cases and $m_h$ becomes larger than 128\,GeV: larger $|c_H|$ leads to smaller (larger) squark (stop) masses. Considering the Higgs boson mass constraint and the sizes of the squark mass and the lightest neutralino mass, it is expected that the regions for $m_{3/2} \lesssim 300$\,-\,350\,TeV can be tested at the high-luminosity LHC~\cite{highlumi}.\footnote{
One may consider the case that the $R$-parity is slightly violated. In this case, the viable 
region is much wider, since the sleptons and squarks can be lighter than ${\chi_1^0}$.
 The region with $m_{3/2}\lesssim 700$\,TeV may be tested at the LHC by searching an $R$-hadron. 
 Notably, the region for even heavier gravitino may be also tested by searching a stable slepton 
 in both LHC and ILC. (The sleptons can be $\mathcal{O}(100)$\, GeV even for $m_{3/2}$ = 1000 \,TeV.)%
%We demand the slepton be heavier than $m_{\chi_1^0}$ in a small region where $m_{\chi_1^0} < 340$ GeV.
}

\paragraph{Regions favored by the thermal leptogenesis}
%Here, we consider the case that the wino-like neutralino is stable. 
%
If the thermal leptogenesis is responsible for the observed baryon asymmetry, 
the mass of the wino-like neutralino needs to be smaller than about 1\,TeV~\cite{Ibe:2011aa} when it is the
stable DM. This is because the neutralino produced from the gravitino decay leads to the over-closure of the universe for $m_{\chi_1^0} \gtrsim 1$\,TeV, if the reheating temperature is higher than $\sim 10^{9}$\,GeV due to the large energy density of the gravitino. 
Note that the reheating temperature higher than about $\sim 10^{9}$\,GeV is required for the successful thermal leptogenesis~\cite{Giudice:2003jh,Buchmuller:2004nz}. The critical value, $m_{\chi_1^0} \simeq 1$\,TeV, corresponds to $m_{3/2} \simeq 320$\,TeV; therefore the regions with $m_{3/2} \sim 300$\,TeV or smaller are especially interesting.

In Fig.~\ref{fig:cont1}, we show the contours of the lightest squark mass (left) and slepton mass (right) as well as $m_h$. 
On the upper (lower) two panels, we take $m_{3/2}=150$ (300)\,TeV. 
%The Higgs boson mass is computed using {\tt SUSYHD 1.0.2}~\cite{Vega:2015fna}. 
%
In the gray regions, the lightest slepton is lighter than 340 GeV or the lightest squark is lighter than 1\,TeV; hence, those regions are likely to be excluded. In the blue shaded region, the slepton is lighter than $\chi_1^0/\chi_1^\pm$ and the LSP. 
In the green (pink) shaded regions, the lightest squark (slepton) is the right-handed down squark (left-handed selectron). [Thus, the right-handed strange squark (the left-handed smuon) is also light.]
In the other regions of the left (right) panels, the right-handed up squark (right-handed selectron) is the lightest squark (slepton). (The right-handed smuon is also light.)
In the wide regions, the lightest squark is lighter than 3\,TeV, which may be discovered at the future LHC experiments. Also, the slepton can be light as 500\,GeV for $m_{3/2}=150$\,TeV.

\paragraph{Muon $g-2$}
Finally, let us briefly comment on regions where the muon $g-2$ anomaly~\cite{Bennett:2006fi, Roberts:2010cj,Hagiwara:2011af}\footnote{
See also \cite{Davier:2010nc} for a standard model prediction of the muon $g-2$.
}
 is explained~via SUSY contributions~\cite{Lopez:1993vi,Chattopadhyay:1995ae,Moroi:1995yh}.
Since the sleptons and bino can be light as $\mathcal{O}(100)$\,GeV and $\simeq 1$\,TeV, respectively for $m_{3/2} \sim 100$\,TeV 
 and $\mu \tan\beta$ is as large as $\sim \mathcal{O}(10^3)$\,TeV, there are regions consistent with the muon $g-2$ experiment.
In Fig.\,~\ref{fig:g-2}, we show the regions consistent with the experimental value of the muon $g-2$ at $1\sigma$ (red), 1.5$\sigma$ (orange) and 2$\sigma$ (yellow) level.
In the left (right) panels, the contours of the lightest squark (slepton) mass are also shown. We take $m_{3/2}=100$ (140) TeV in the upper (lower) two panels. We include leading two-loop corrections: the correction to the muon Yukawa coupling~\cite{Marchetti:2008hw} and the logarithmic QED correction~\cite{Degrassi:1998es}. 
These regions are explored in details in Ref.~\cite{Yin:2016shg}, where further information such as the bottom-tau Yukawa unification and bottom-tau-top Yukawa unification can be found.\footnote{
SUSY models beyond the MSSM explaining the muon $g-2$ are shown in Refs.~\cite{Endo:2011mc, Endo:2011xq, Endo:2011gy, Moroi:2011aa, Nakayama:2012zc, Sato:2012bf, Shimizu:2015ara} 
}

\subsection{Case of non-vanishing squark and slepton masses}

%%%%%%%%%%%%
\begin{figure}[!t]
\begin{center}
\includegraphics[bb=100 150 580 700, scale=0.4]{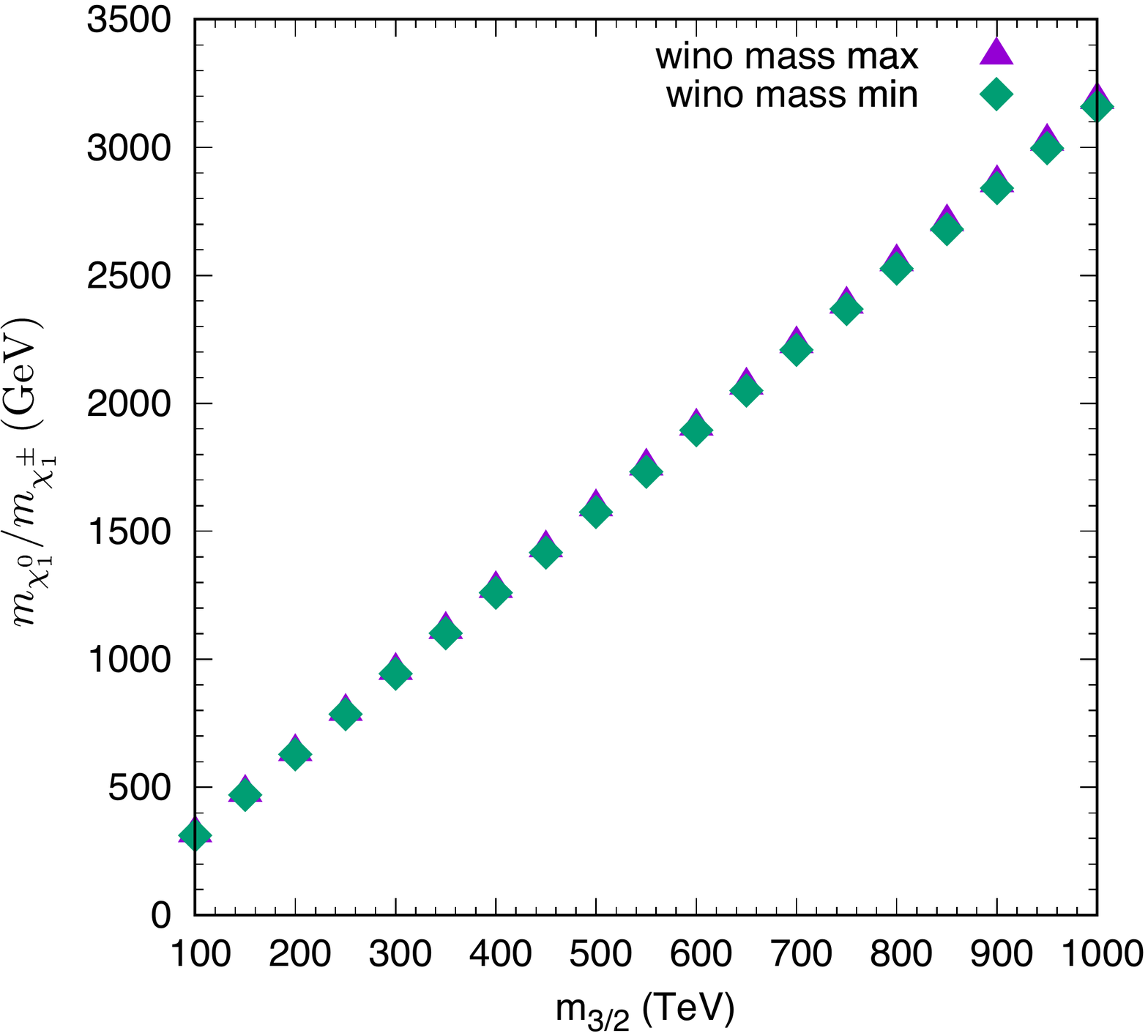}
\includegraphics[bb=0 150 500 650, scale=0.4]{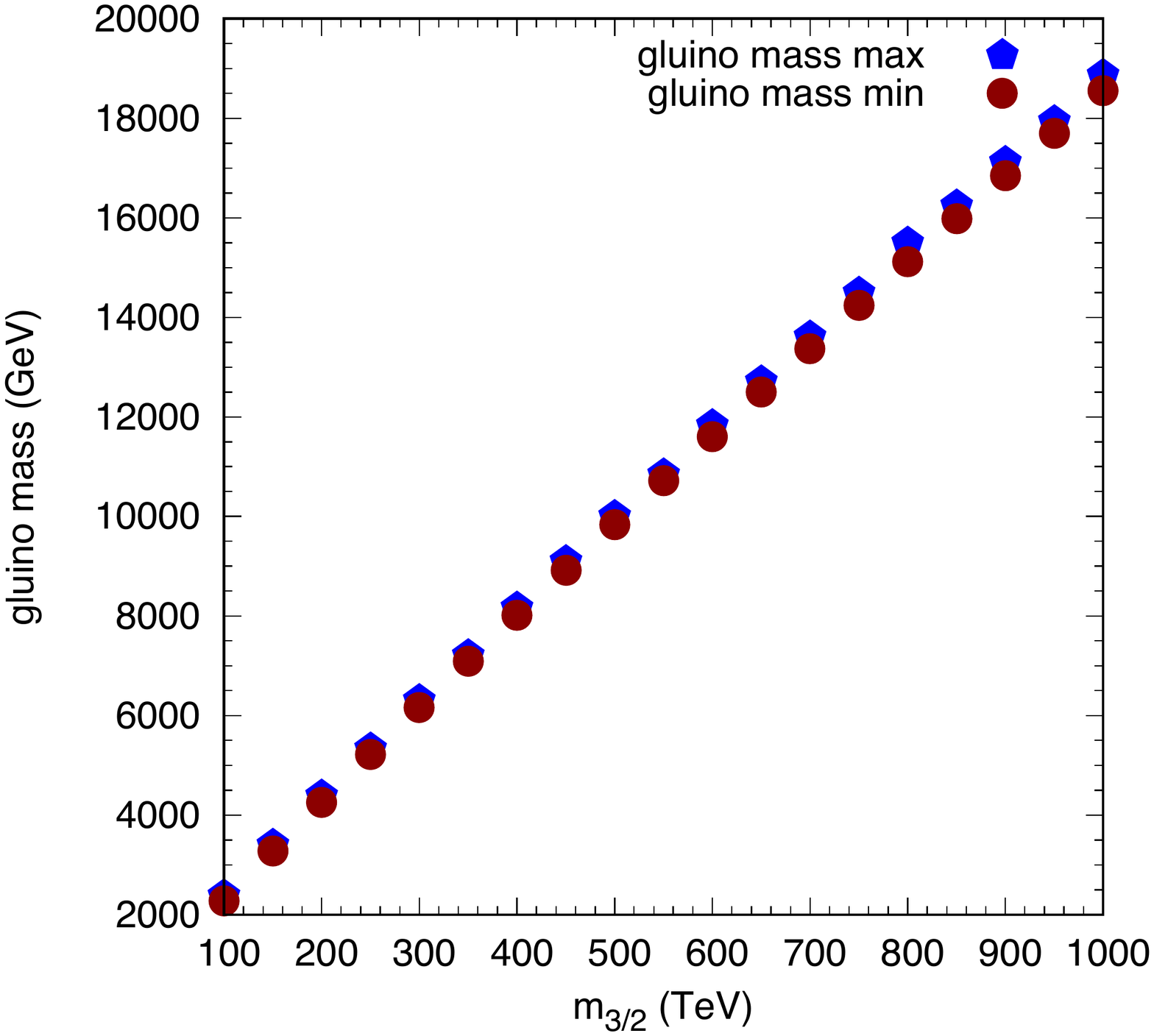}
\caption{
The masses of the lightest chargino/neutralino and the gluino with $m_0$. The triangles and pentagons (squares and circles) show maximum (minimum) values. 
}
\label{fig:3}
\end{center}
\end{figure}
%%%%%%%%%

%%%%%%%%%%%%
\begin{figure}[!t]
\begin{center}
\includegraphics[bb=100 170 610 700, scale=0.38]{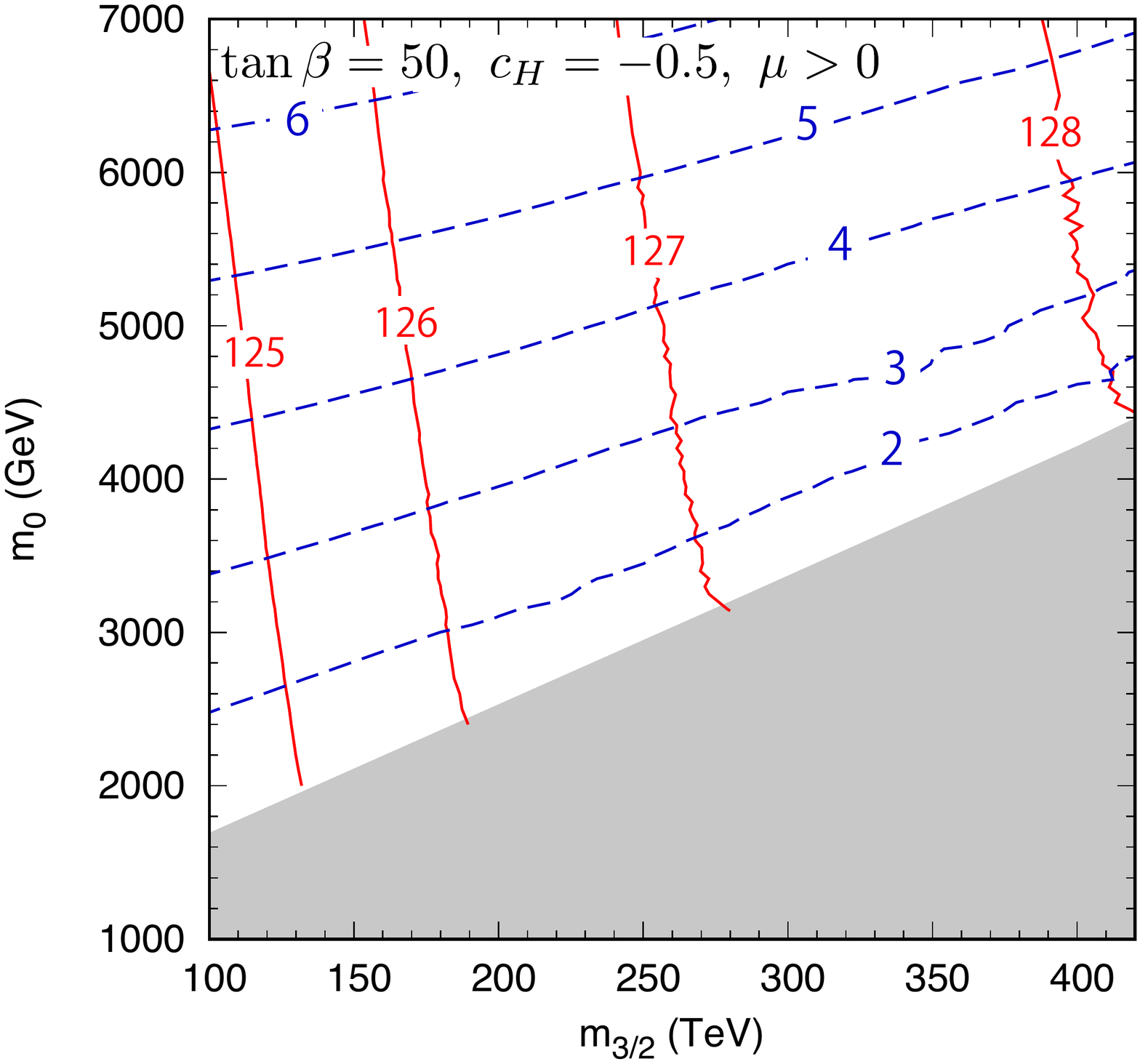}
\includegraphics[bb=0 170 520 700, scale=0.38]{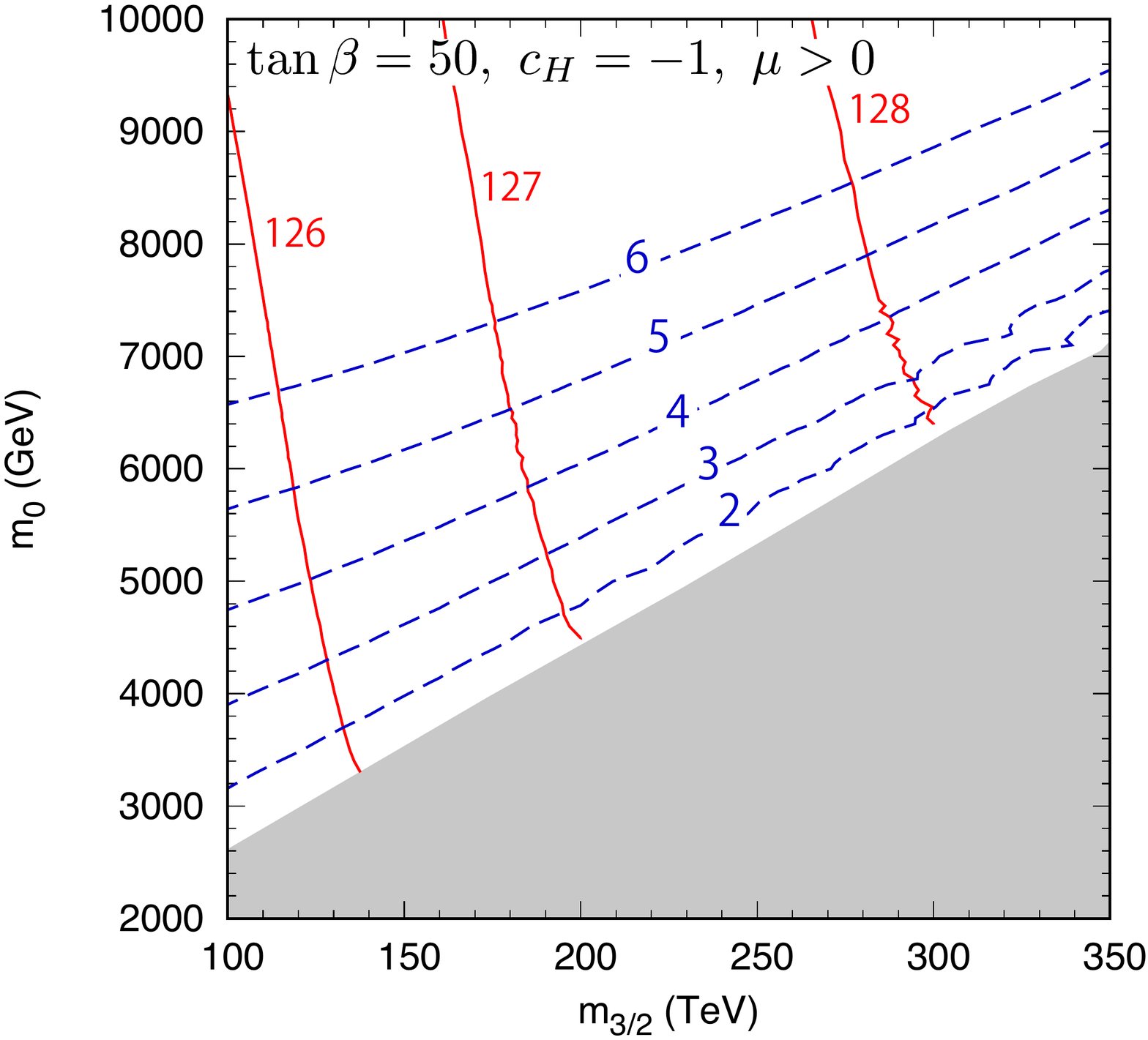}
\caption{
The contours of $m_h$ (GeV) and $m_{\tilde q}$ (TeV) for $c_H=-0.5$ and $-1$. The red solid (blue dashed) line shows $m_h$ $(m_{\tilde q})$.
Here, $m_{\tilde q}$ is a mass of the lightest squark.
%In the red (green) region, the muon $g-2$ is explained at 1$\sigma$ (2$\sigma$) level.
}
\label{fig:squark_m0}
\end{center}
\end{figure}
%%%%%%%%%

Since Yukawa and gauge couplings break $E_7$ explicitly, there might be small SUSY-breaking soft masses for the squarks and sleptons. In this subsection, we consider the case that squark and sleptons have a common soft mass $m_0$ of $\mathcal{O}(1)$\,TeV 
in addition to the contribution from anomaly mediation in Eq.\,(\ref{eq:amsb_scalar}). Notice that $m_0$ breaks the $E_7$ symmetry explicitly.
The model parameters in this setup are 
\begin{eqnarray}
m_0, \ m_{3/2}, \  \tan\beta   ,\ c_H, \ {\rm sign}(\mu).
\end{eqnarray}

The masses of the lightest chargino and the gluino are shown in Fig.~\ref{fig:3}. We take $\ {\rm sign}(\mu)=+$. 
The maximum (minimum) values are denoted as triangles and pentagons (squares and circles).
The lightest chargino (neutralino) is almost pure wino as in the case of Sec.~\ref{sec:vanish}. 
We scan the parameter space over the ranges
\begin{eqnarray}
40<\tan\beta<60, \ \ 1000\,{\rm GeV} + m_{3/2}/100 < m_0 < 2(1000\,{\rm GeV} + m_{3/2}/100),
\end{eqnarray}
with the fixed $c_H$, $c_H=-1$. The result is very similar to the one in Fig.~\ref{fig:1}, where there is an approximate one-to-one correspondence between the chargino mass and the gravitino mass.

\begin{table*}[!t]
\caption{\small Mass spectra in sample points.
}
\label{tab:sample}
%\begin{table}[]
\begin{center}
\begin{tabular}{|c||c|c|c|c|}
\hline
Parameters & Point {\bf I} & Point {\bf II}  & Point {\bf III}  & Point {\bf IV}\\
\hline
$m_{3/2} $ (TeV) & 320  & 140  & 250 & 100 \\
$c_H$  & $-0.25$  & $-0.06$  & $-0.2$ & -1 \\
$\tan\beta$  & 50  & 47  &  62 & 50 \\
$m_0$\,(GeV) & 0 & 0 & 0 & 3000\\
sign($\mu$) &  $+$ &  $+$ & $-$ & $+$\\
\hline
%\hline
%
Particles & Mass (GeV) & Mass (GeV)& Mass (GeV) & Mass (GeV) \\
\hline
$\tilde{g}$ & 6350 & 2960 & 5060  & 2330 \\
$\tilde{q}$ & 2340\,-\,5470 & 2130\,-\,2410 & 2990\,-\,4430 &  1760\,-\,3490\\
$\tilde{t}_{2,1}$ (TeV) & 62, 62 & 15, 14 & 44, 41 & 39, 37\\
$\tilde{b}_{2,1}$ (TeV) & 65, 64 & 16, 15 & 42, 39& 39, 38\\
$\tilde \chi_1^0$/$\tilde{\chi}_{1}^\pm$ & 993 & 441  & 790 & 323\\
$\tilde{\chi}_2^0$ & 2990 & 1290 & 2340 & 934 \\
$\tilde{e}_{L, R}$ & 4200, 3310 & 700, 662  & 3150, 1720 & 4230, 3530\\
$\tilde{\mu}_{L, R}$ & 4480, 3980 & 757, 779  & 4530, 4900 & 4420, 3980\\
$\tilde{\tau}_{2,1}$ (TeV) & 60, 43&  13, 8.9 & 52, 37 & 38, 27\\
$H^\pm$\,(TeV) & 47 & 12 & 25 & 22\\
$h_{\rm SM\mathchar`-like}$ & 126.6 &  122.3  & 125.5 & 125.2\\
\hline
$\mu$ (TeV) & 140  & 30  & -98 & 88\\
\hline
%$\Delta$ & 126 & 125 & 127 & 126\\
%gravitino & 19.2 (eV) & 4.81 (eV) & 3.85 (eV) & 3.85 (eV) \\
%\hline
%mass difference & (GeV) & (GeV) & (GeV) \\
%\hline
%$m_{\tilde{g}} - m_{\tilde{\chi}_1^0}$ & 198 & 256 & XXX \\
%$m_{\tilde{t}_1} - m_{\tilde{\chi}_1^0}$ & 67 & 26.9 & XXX \\
%$m_{\tilde{u}_L} - m_{\tilde{\chi}_1^0}$ & 122 & 133 & XXX \\
%\hline
\end{tabular}
%\end{table}
\end{center}
\end{table*}

In Fig.\,\ref{fig:squark_m0}, we show the contours of the squark mass and $m_h$ on $m_{3/2}$-$m_0$ plane for
 $c_H=-0.5$ (left) and -1 (right). We set $\tan\beta=50$. We focus on the viable parameter space, where $122 \le m_h \le 128$, corresponding to the gravitino mass up to 400 (300) TeV for $c_H=-0.5 (1)$.
If the universal scalar mass at $M_{\rm inp}$ is smaller than 3\,-\,4 TeV, the lightest squark is lighter than about 3.5\,TeV, which may be accessible to the future LHC experiments.

\subsection{Mass spectra}

We show mass spectra at some example points in the viable parameter space of the model (Table~\ref{tab:sample}).
%The Higgs boson mass is computed using {\tt SUSYHD 1.0.2}~\cite{Vega:2015fna}. 
On the points {\bf I}-{\bf III}, the masses of the squarks and sleptons vanish at $M_{\rm inp}$, while on the point {\bf IV}, we introduce the small universal mass $m_0$. On the point {\bf III}, ${\rm sign}(\mu)=-$; therefore, $\tan\beta$ is large as $\sim 60$.
On the point {\bf II}, the smuons are light and the SUSY contribution to the muon $g-2$ is $1.4 \times 10^{-9}$. 
On that point, the calculated Higgs boson mass using {\tt SUSYHD} is $\simeq 122$\,GeV. However, the stop mass is as large as 15\,TeV, and the computed Higgs mass using {\tt FeynHiggs 2.12.0}~\cite{feynhiggs, feynhiggs2, feynhiggs3, feynhiggs4, feynhiggs5} is larger than 125\,GeV; therefore, the point could be consistent with the observed Higgs boson mass. On the listed points, the mass of the lightest squark is smaller than 3\,TeV, which may be tested at the LHC Run-2 or at the high luminosity LHC.

\section{Conclusion}
The $E_7/SU(5)\times U(1)^3$ non-linear sigma model is very  attractive, since it may provide us an intriguing answer to one of the fundamental questions, why we have three families of quarks and leptons. In the $E_7$ NLS model, the masses of squarks and sleptons vanish at the tree level while the Higgs doublets have soft SUSY breaking masses of the order of the gravitino mass. 

In this paper, we have shown that the $E_7/SU(5)\times U(1)^3$ NLS model is consistent with all observations if one adopts the pure gravity mediation or minimal split SUSY, where the gaugino masses arise only from anomaly mediation. The tachyonic slepton problem in anomaly mediation is solved due to the renormalization group running effects from the negative Higgs soft mass squares. 
We have shown that if the observed baryon asymmetry is explained by the thermal leptogenesis, the squarks are lighter than 2-3 TeV in a wide range of the viable parameter space, and we expect them to be discovered at the LHC Run-2 or at the high luminosity running. Moreover, the sleptons may be as light as O(100)\,GeV, giving rise to a possibility for explaining the muon $g-2$ anomaly. The sleptons are also interesting target at the LHC and at ILC. 

Although we have concentrated on the case that the $R$-parity is conserved, one can also consider the small $R$-parity violation. In this case, it is easy to imagine that the viable parameter region becomes wider, since the squarks and sleptons can be lighter than the lightest neutralino. Note that the testability of the model is also enhanced: the stable squark can be checked by searching an $R$-hadron for heavy gravitino of $\sim$\,700 \,TeV; the stable slepton can be as light as $\mathcal{O}$(100)\, GeV even for $m_{3/2}$ = 1000 \,TeV.

Finally, let us comment on the SUSY CP-problem for $m_{3/2}=100$\,-\,$300$ \,TeV, 
where the selectron, bino and first/second generation squarks are as light as a few TeV. 
In this case, the constraints from the electron and neutron electric dipole moment (EDM) are severe. 
For instance, in the base that the $\mu$-term and gravitino mass are real, the argument of the Higgs $B$-term needs to be as small as $10^{-3}$\,-\,$10^{-4}$ to avoid the constraint from the electron EDM~\cite{Baron:2013eja}. It is, however, remarkable that we need  CP violation only in the $E_7$-breaking Yukawa couplings, so far.
Thus, it is very interesting to consider that violations of two independent symmetries, CP and $E_7$, arise from
 an underlying common physics.

\section*{Acknowledgments}
This work is supported by JSPS KAKENHI Grant Numbers 
JP26104009 (T.T.Y), JP26287039 (T.T.Y.), JP16H02176 (T.T.Y),
JP15H05889 (N.Y.) and JP15K21733 (N.Y.);
and by World Premier International Research Center Initiative (WPI Initiative), MEXT, Japan (T.T.Y.).

\appendix

\section{Vanishing soft mass for NG multiplet with direct coupling to the SUSY breaking field}
To explicitly see that the soft mass of the NG multiplet  vanishes even if there is a direct coupling to the SUSY breaking field,  
let us consider the following K{\" a}hler potential and superpotential:
\begin{eqnarray}
K = f(x)(1+c Z^\dag Z) + Z^\dag Z, \ \ W= W(Z) + \mathcal{C},
\end{eqnarray}
where $x=\phi^\dag \phi + S + S^\dag$ and $\mathcal{C}$ is a constant term; $S$ is required for the $E_7$ invariance of the Lagrangian.   
We take unit of $M_P=1$.
\begin{eqnarray}
K_{\phi\phi} &=& (f' + |\phi|^2 f'')(1+c |Z|^2)  \nonumber \\
K_{SS} &=& f'' (1+c |Z|^2) \nonumber \\
K_{ZZ} &=& 1 + cf \nonumber  \\ 
K_{S\phi} &=& \phi^\dag f'' (1+c |Z|^2) \nonumber \\
K_{Z\phi} &=& \phi^\dag f' (c Z) \nonumber \\
K_{ZS} &=& f' (c Z),
\end{eqnarray}
where $K_{ij} = \frac{\partial^2 K}{\partial q_i^\dag q_j}$. We see that 
\begin{eqnarray}
{\rm det}(K) &=& - (1+c |Z|^2) f' [c^2 |Z|^2 f'^2 - (1+c |Z|^2)(1+c f) f'' ] \nonumber \\
&=& F(x, |Z|^2),
\end{eqnarray}
where the dependence on $|\phi|^2$ arises only though $x$.

First consider the simple case: $\left<Z\right>=0$,
\begin{eqnarray}
K_{\phi\phi} &=& f' + |\phi|^2 f'' \,, \nonumber \\
K_{SS} &=& f'' \,,  \nonumber \\
K_{ZZ} &=& 1 + cf \,, \nonumber \\ 
K_{S\phi} &=& \phi^\dag f''  \,, \nonumber \\
K_{Z\phi} &=& 0 \,, \nonumber \\
K_{ZS} &=&  0 \,.
\end{eqnarray}
The inverse matrix is almost same as in the case of Sec.~2; thus, the argument in Sec.~2 is still valid, even if there is an additional term:
\begin{eqnarray}
V \ni e^{K} \left|\frac{\partial W}{\partial Z}\right|^2 \frac{1}{1+c f} = G_1(x, |Z|^2).
\end{eqnarray}
So far, $\frac{\partial^2 V}{\partial \phi^\dag \partial \phi}$ vanishes at the minimum.

Next we consider the case $\left<Z\right>\neq 0$. The inverse matrix is given by
\begin{eqnarray}
{\rm det}(K)\, K^{-1}_{\phi \phi} &=& -c^2 |Z|^2 f'^2 + (1 + c|Z|^2) (1 + c f) f''  \,, \nonumber \\ 
{\rm det}(K)\, K^{-1}_{S S} &=& -c^2 |Z|^2 |\phi^2| f'^2 + (1+c|Z|^2)(1+cf)(f'+|\phi|^2 f'')  \,, \nonumber \\ 
{\rm det}(K)\, K^{-1}_{Z Z} &=& (1+c |Z|^2) f' f''  \,, \nonumber \\ 
{\rm det}(K)\, K^{-1}_{\phi S} &=&  \phi \left[ c^2 |Z|^2 f'^2 - (1+c|Z|^2) (1+cf) f'' \right]  \,, \nonumber \\ 
{\rm det}(K)\, K^{-1}_{\phi Z} &=& 0 \,,  \nonumber \\ 
{\rm det}(K)\, K^{-1}_{S Z} &=& -c (1+c |Z|^2) Z^\dag f'^2  .
\end{eqnarray}
Therefore, only
\begin{eqnarray}
V \ni e^{K} |W|^2 \left[ \left|\frac{\partial K}{\partial \phi} \right|^2 K^{-1}_{\phi\phi} 
+ \left|\frac{\partial K}{\partial S}\right|^2 K^{-1}_{S S}
+\frac{\partial K}{\partial S} \frac{\partial K}{\partial \phi^\dag} K^{-1}_{S \phi} + h.c.
\right] = G_2 (x, |Z|^2)
\end{eqnarray}
is relevant as in Eq.~(\ref{eq:gs_mass}). This $G_2$ gives the vanishing soft mass to the NG multiplet.

\end{document}